\documentstyle[aabib,psfig,epsf]{l-aa}

\newcommand{\msun}{\mbox{${\rm M}_\odot$}}

\newcommand{\rsun}{\mbox{${\rm R}_\odot$}}
\newcommand{\Kms}{\mbox{${\rm km~s}^{-1}$}}
\newcommand{\Zytkow}{$\dot{\mbox{Z}}$ytkow}
\newcommand{\TZO}{$\mbox{T}\dot{\mbox{Z}}$\mbox{O}}
\newcommand{\vsun}{\mbox{${\rm v}_\odot$}}

\begin{document}
\thesaurus{03.13.4;
           05.03.1;
           08.05.3;
           08.02.7;
           10.07.2}

\title{Star Cluster Ecology I}
\subtitle{A Cluster Core with Encounters between Single Stars}
\author{Simon F.\ Portegies Zwart\/$^{1, 2}$, P.\ Hut\/$^3$ \&\ F.\ Verbunt\/$^1$}
\offprints{Simon Portegies Zwart}

\institute{$^1$ Astronomical Institute, Postbus 80000,
                3508 TA Utrecht, The Netherlands \\
           $^2$ Astronomical Institute {\em Anton Pannekoek}, 
                Kruislaan 403, 1098 SJ Amsterdam, The Netherlands \\
           $^3$ Institute for Advanced Study, 
                Princeton NJ, USA}

\date{received; accepted:}
\maketitle
\markboth{S.F.\ Portegies Zwart, P.\ Hut \&\ F. Verbunt}
         {Star Cluster Ecology I}

\begin{abstract}
Star clusters with a high central density contain an ecological
network of evolving binaries, affected by interactions with passing
stars, while in turn affecting the energy budget of the cluster as a
whole by giving off binding energy.
This is the first paper in a series aimed at providing the tools for
increasingly realistic simulations of these ecological networks.  

Here we model the core of a globular star cluster.  The two main
approximations are: a density of stars constant in space and time, and
a purely single star population in which collisions between
the evolving stars are modeled.
In future papers in this series, we will relax these
crude approximations.  Here, however, they serve to set the stage
before proceeding to the additional complexity of
binary star interactions, in paper II, and background dynamical
evolution, in later papers.

\end{abstract}

\section{Introduction}
In dense stellar systems, such as open and globular clusters and
galactic nuclei, encounters between individual stars and binaries can
affect the dynamical evolution of the system as a whole on a time
scale comparable to, or even shorter than, a Hubble time.  In
order to reach a detailed theoretical understanding of such systems,
the following three steps are necessary.

First, we need to understand the basic mechanism of the dynamical
evolution, in the limit of a point-mass approximation for the stars.
Second, effects of dynamical encounters on the internal evolution of
single stars and binaries has to be taken into account.  Third, we
have to model the feedback of these internal changes onto the
dynamical evolution of the whole system.  Let us briefly review each
step.

Great progress has been made with the first step, modeling the
dynamical evolution of point-mass systems.  In the seventies, the
processes of core collapse and mass segregation were studied with the
use of various types of Fokker-Planck approximations.  In the
eighties, these simulations were extended successfully beyond core
collapse, and various studies were made of the phenomenon of
gravothermal oscillations, ubiquitous in the post-collapse phase. 
Of these models a few even include mass loss 
due to the evolution of the stars (\cite{cw90}).
In the nineties, we are finally beginning to switch over from
Fokker-Planck approximations to much more detailed and realistic
$N$-body simulations.  In 1995, the construction of the GRAPE-4, a
special-purpose machine with Teraflops speed, has made a
$32,000$--body simulation feasible, providing the first direct
evidence of gravothermal oscillations (Makino, 1996a,b).  Extending
these simulations to the full realm of globular clusters
($N=10^5\sim10^6$)
will require Petaflops speed, something that could be realized by
future special-purpose machines in the GRAPE series by as early as the
year 2000.

While the point-mass approximation provides a good qualitative guide
for the construction of dynamical models of dense stellar systems,
this approximation quickly breaks down when we require quantitatively
accurate results.  The second step attempts to model the effects of
close encounters.  A number of different investigations have
estimated the rate at which physical collisions have take place, under
various circumstances 
(\cite{hd76}, \cite{vm88}, \cite{dr92} and \cite{db95}). 
However, little progress has been made so far in following the changes
induced
in the stellar population, beyond enumerating the number of mergers.
In the simulations presented below, collisions are modeled in an
evolving population of single stars in a high-density stellar
environment.
Paper II in this series will extent our treatment to follow the induced changes
in binary systems, both on the level of changes in orbital parameters
as well as in the internal structure of the stars.

However, such investigations are only a start, and cannot lead to a
quantitative modeling of dense stellar systems, since they are not yet
self-consistent.  What is needed in addition is a treatment of the
feedback mechanism, from the changes in single stars and binaries back
to the overall dynamics of the system. 
This third step is being pioneered for open clusters by \cite*{aar96}. 
The current series of papers aims to provide self-consistent models of
this type, by coupling relatively crude stellar evolution recipes,
documented and tested in papers I and II, to a fully dynamical
$N$-body system.

This paper is organized as follows.  Our approach to the study of the
ecology of star clusters is summarized in somewhat more detail in \S 2.
The next section, \S 3, describes our simulation techniques, and the
various approximations involved.  In \S4, we present the result of a
simulation starting a single model, with a minimum of free parameters.
The results of a more realistic core model run are presented in \S5.
\S6 sums up.

\section{Star Cluster Ecology}

Stellar evolution plays a role in star cluster evolution similar to
the role played by nuclear physics in stellar evolution.  In both
cases, the microphysical processes play a crucial role in the
mechanism of energy generation in the central parts of the system
under consideration, a mechanism that tries to balance the energy
losses at the outskirts (tidal radius and photosphere, respectively).

In the next two subsections we look separately at the different forms
of physics input necessary to follow the evolution of a star cluster.
The third subsection then discusses their interconnection.  For
future reference, initial conditions are discussed in the fourth
subsection, while the last subsection provides a brief outline of the
series of papers of which this one is the first.

\subsection{Stellar Dynamics Simulations}
Great progress has been made in the study of star cluster dynamics,
using various approximate methods in which the stars have been treated
like a form of fluid, either three-dimensional as in conducting gas
sphere models, or six-dimensional as in Fokker-Planck models.  In both
cases, the main effect of encounters has been taken into account by a
form of effective two-body relaxation.  We refer to \cite*{hmg+92}
for a review of these methods.

Unfortunately, both methods have two intrinsic handicaps that make
them unsuitable for a detailed quantitative modeling of the evolution
of a globular cluster past core collapse.  First, they are not set up
to deal with the separate evolution of internal and external degrees
of freedom of the binaries that play a central role in the energy
generation processes in the cluster.  

The second problem stems from an introduction of a mass spectrum, as
well as a distinction between stars of different radii, such as
dwarfs, main-sequence stars, and giants.  The root of the problem here
is that a gas sphere or Fokker-Planck approach does not follow
individual stars, but rather distribution functions.  When the number
of independent parameters characterizing the distribution functions
becomes too large, there will be less than one star left in a typical
cell in parameter space --- something that clearly invalidates the
statistical hypothesis on which these methods are based.

The only solution is to drop the statistical assumption, and to revert
to a star-by-star modeling of a globular cluster, through direct
$N$-body calculations.  The draw back of such an approach has long
been the prohibitive calculational costs involved, and until recently
typical production runs only included a few thousand stars.  To
extend such numbers to include several hundred thousand stars,
characteristic of realistic globular clusters,
requires an increase of two orders of magnitude in star number, or a
factor million in computational cost, from Gigaflops days to Petaflops
days (\cite{hmm88}).

Recently, the number of stars modeled in direct $N$-body calculations
has been increased significantly, to $N=32,000$, using the GRAPE-4, a
form of special-purpose hardware developed by a group of
astrophysicists at Tokyo University, running at a speed of 1 Tflops
(Makino 1996a).  The first scientific results of the GRAPE-4,
including the first convincing evidence of gravothermal oscillations
in $N$-body simulations, predicted by \cite*{sb83},
have been presented by Makino (1996ab).
\nocite{mak96a}\nocite{mak96b}

The next, and definitive step that will enable any globular cluster to
be modeled realistically might take place as early as the year 2000.
If funding can be found, there is no technological obstacle standing
in the way of a speedup of the current GRAPE-4 machine by a factor of
a thousand, during the next five years.  Most of this speed-up will
come from further miniaturization, allowing a larger number of gates
to be mounted on a single chip, and allowing a higher clock speed as
well.  A Petaflops machine by the year 2000, allowing simulations of
core collapse and post-collapse evolution with up to $10^6$ particles,
is thus a realistic goal.

\subsection{Stellar Evolution Population Synthesis}
The first serious attempts to understand and simulate the 
evolution of close binaries were made in the mid thirties (\cite{hh37})
and late fifties by \cite*{cra55}, \cite*{kop56} and \cite*{hua56} 
followed by \cite*{mor60} and the standard work in binary 
evolution from \cite*{kw67}.
Synthesis of complete populations of single stars became popular in the
mid seventies when \cite*{tg76} simulated the giant-branch luminosity
functions for giant elliptical galaxies.
However, it is only recently that detailed studies simulate complete
populations of close binaries starting with \cite*{dc87} who tried to
understand the evolutionary sequence of radio pulsars and the presence
of an asymmetry in the velocity distribution of single radio pulsars.
In later papers, similar evolutionary scenarios for the formation of
binary neutron stars were studied in more detail 
(\cite{ty93}, \cite{lpp+95}, \cite{pzs96} and \cite{lpp96}), 
for high mass X-ray binaries and the supernova rate in the galaxy 
(\cite{tyi92}, \cite{lip94}, \cite{ds95}, \cite{pzv96}) 
and for lower mass systems with a neutron star 
(\cite{wk93}, \cite{pm93}) or a white dwarf (\cite{koo90}, \cite{kr92})
as the accreting object. 
The first couragous attempt to combine stellar and binary evolution
within
the collisional evironment of a globular cluster
was performed by \cite*{sut75}, followed more recently by 
\cite*{dr92}, \cite*{sp93}, \cite*{leo95}, \cite*{db95} and
\cite*{dav95}.

\subsection{Ecological Networks}

Purely stellar-dynamical calculations often rely on rather severe
approximations, such as a representation of stars by equal-mass
point-masses.  And there is a good reason for doing so, since any
single deviation from that simple recipe requires other deviations as
well.  Let us look at one example.

As soon as we introduce a mass spectrum in a star cluster simulation,
we will see that the heavier stars start sinking toward the center, on
the dynamical friction time scale, shorter than the two-body
relaxation time by a factor proportional to the mass ratio of individual
heavy stars with respect to that of typical stellar masses.
The reason is that relaxation tends toward equipartition
of energy, which implies that heavier stars will move more slowly and
therefore gather at the bottom of the cluster potential well.

If stars would live forever, there would be a large overconcentration
of heavy stars in the core of a star cluster.  However, in reality
there is an important counter-effect: heavy stars burn up much faster
than lighter ones.  They may or may not leave degenerate remnants, that
may or may not be heavier than the average stellar mass in the cluster
(a quantity that decreases in time).  Clearly, it would be
grossly unrealistic to introduce a mass spectrum without removing most
of the mass of the heaviest stars on the time scale of their evolution
off the main sequence and across the giant branches.

Another reason for introducing finite life times for stars comes from
abandoning the very restrictive point mass model.  As soon
as we do that, giving our stars a finite radius will give rise to
stellar collisions.  The heavier stars produced in the collision of two
turn-off stars, for example, will burn up on a time scale an order of
magnitude smaller than the age of the cluster.  Again, we
have to take this into account to be consistent, especially since the
merger products themselves are prime candidates for further merging
collisions.

The need to let many stars shed most of their mass, together with the
fact that most of the energy in a globular cluster is locked up in
binaries, poses a formidable consistency problem.  Since binaries play
a central role in cluster dynamics, consistency requires that we
follow their complex stellar evolution, which involves mass overflow
(which can be stable or unstable, and can take place on dynamical or
thermal or nuclear time scales) and the possibility of a phase of
common-envelope evolution.  On top of all that, we will have to find
simple recipes for the hydrodynamic effects occurring in three-body and
four-body reactions, and in occasional $N>4$ reactions, which are
bound to occur in dense cluster centers.

To sum up: there does not seem to be a half-way stopping point, at
which we can expect to carry out consistent cluster evolution
simulations.  Either we study the interesting but unrealistic
mathematical-physics problem of an equal-mass point particle model, or
we opt for a realistic model with some set of stellar-evolution
recipes.  The main question here is: what is the simplest set that is
still
consistent?

\subsection{Initial Conditions}

In most stellar dynamics simulations of star clusters, the Plummer
model is used as a standard model to specify the initial conditions for
the distribution of the point particles.  While not very realistic,
this choice has had the advantage of making comparisons between
different runs, as well as between different approaches, relatively
straightforward.  Of course, when attempts are made to model
particular star clusters, other models have often replaced the Plummer
model as a starting point.  King models, for example, are already more
realistic in that they provide a form of spatial cut-off that can be
interpreted as a tidal radius.

For similar reasons, we will use a standard model for our simulations
that combine stellar dynamics and stellar evolution.  In most cases,
the use of our standard model will be mostly for illustrative reasons,
to provide a gauge for comparison between our
various results, as well as between our results and that of others.
For historical reasons, we choose our standard model to be based on a
Plummer model
for the macroscopic initial star distribution, and a Salpeter model
for the initial mass function.

An additional advantage of these simple choices is that they limit the
number of free parameters.  The Plummer model, for example, contains
only one free parameter, $N$, the number of stars in the system (apart
from a choice of mass and length scales, that are irrelevant in the
point particle
case).  In contrast to the Plummer model, our standard model can be
expected to form a multi-parameter family.  As soon as we abandon the
point-mass approach, we have to deal with microscopic as well as
macroscopic mass and length scales.

Of these various scales, the macroscopic quantities can be chosen
independently, while the microscopic ones can be fixed,
statistically, by specifying a mass distribution together with
appropriate cut-off masses at the high and low end.  In general, an
arbitrary
functional form for the mass distribution function can lead to an
arbitrarily large number of parameters.  Interestingly, our standard
model definition allows us to limit the total number of free
parameters to three.

Starting from the macroscopic side, we can take the total mass $M$ and
the half-mass radius $r_h$ of the Plummer model as our first two free
parameters. With a Salpeter choice of powerlaw distribution function, 
the third free parameter can be chosen in the form of the lower mass
cut-off $m_{-}$.  The higher-mass cut-off $m_{+}$ could be
specified independently, but this is not strictly necessary: since the
Salpeter distribution function converges at the high-mass end, we can
simply parcel out the total mass $M$ over different stellar
masses, between star masses of $m = m_{-}$ and $m_{+} = \infty$, and we
will naturally be left with a single most massive star.  This
procedure is not unrealistic: nature probably limits the number of
high-mass stars in medium-size galactic clusters in a similar way.

In fact, we can go even further, and make the following somewhat
arbitrary but natural choices: $r_h = 10$~pc, $m_{-} = 0.1 \msun$.
This leaves only the total mass $M$ to be specified, or equivalently,
the total number of stars $N$.  For future convenience, we will refer
to this `most standard' model as our reference model.  For systems
with a few thousand stars, we are dealing
with a typical open cluster, with velocity dispersions of order 1
km/s, while for a few hundred thousand stars, we have a reasonable
approximation to a globular cluster, for which typical stellar
velocities are an order of magnitude higher.

In addition to this standard model, the various papers in this series
will also contain the results of more realistic models.  However, we
will typically provide at least one run from a standard model, in
order to provide comparison material for the more detailed models.

\subsection{Stepping Stones}
In the current series of papers, our goal is to provide a series of
ecological simulations, based on a flexible stellar dynamics code
coupled to a comprehensive set of stellar evolution modules.  These
modules in turn are based on recipes that govern the behavior of both
single star and binary star evolution, as well as interactions between
larger numbers of stars.

In order to present results that can be reproduced and critically
assessed by other groups, we clearly document
the recipes used, as well as their coupling to the dynamics. 
With this aim, we give a detailed
description of
our approach in the first few papers in this series, which will form
stepping stones towards a full-fledged ecological star cluster
evolution code.

The present paper starts off with rather extreme approximations for
the stellar dynamics, as well as the stellar evolution parts of our
simulations.  With respect to the former, we start with a laboratory-type
situation, in which we consider a homogeneous distribution of stars,
kept constant in time.  With respect to the latter, we consider a
population of single stars only.  Paper II will relax the second
assumption, by introducing a population of primordial binaries, and
allowing the formation of new binaries as well.  Later papers will
subsequently relax the former assumption, with the ultimate goal of
using a self-consistent $N$-body code.

\section{A Static Homogeneous Environment with Single Stars}
\subsection{Initial Conditions}
In the present paper, we keep the dynamical environment as simple as
possible, in order to focus on the stellar evolution recipes, that are
introduced here and used in subsequent papers as well.  The stellar
distribution is take to be in thermal equilibrium, with a density that
is constant in space and time.  In addition, an additional
simplification is obtained by excluding any primordial binaries, and
ignoring binary formation channels.  Within this setting, random
encounters between single stars will lead to collisions resulting in
the formation of merger products, the evolution of which can then be
followed along with the evolution of the original single stars.

\subsubsection{Initial Mass Function}

While our main aim is to set-up and clarify our stellar evolution
recipes, we present two calculations that could be interpreted as
having a limited astrophysical interpretation, one for the core of an
$\omega$-Centauri-like cluster (\S4), and one for the core of an
M-15-like cluster (\S5).
Our choice of constant density implies that we can only hope to model
the history of a cluster core, not that of a cluster as a whole.  To
specify the mass distribution, we first take our standard choice: a
Salpeter initial mass function (\S 4), which we will use to model a
relatively unevolved core.  Our second choice will be a much more flat
distribution, which is more appropriate for a high-density
post-collapse cluster core (\S 5).

\subsubsection{Mass and Number Densities}

If we specify the mass density $\rho$ for the stars in our cluster
case, we can use the mass function to
determine the number density $n=n(\rho)$.  In a homogeneous medium the
relation is linear, and for the simplest case of a powerlaw mass
function $f(m)\propto m^{-\alpha}$, we find
\begin{equation}
  {n\over\rho} = {{\alpha-2}\over{\alpha-1}}
                 {{m_-^{1-\alpha} - m_+^{1-\alpha}}
                \over{m_-^{2-\alpha} - m_+^{2-\alpha}}},
\end{equation}
where $m_-$ and $m_+$ are the lower and upper mass cut-offs,
respectively.
For the example of an initial Salpeter mass function, $\alpha = 2.35$,
we find
\begin{equation}
  {n\over\rho} = {0.26\over m_-},
\end{equation}
when we neglect the fact that the upper mass cut-off is finite.
The inverse quantity $m_{av}$ is the average stellar mass:
\begin{equation}
  {m_{av}} = {3.9 m_-} = 0.39 \msun 
\end{equation}
in our standard case where we take a lower cut-off mass of $0.1\msun$.
The median mass $m_{med}$ for a Salpeter distribution is
\begin{equation}
  {m_{med}} = 2^{1\over{\alpha-1}} m_{-} = 0.17 \msun 
\end{equation}
which means that in our standard population most stars have a mass
well below $0.2 \msun$.

Even in our simple case of a homogeneous system, the linear
relationship $n(\rho)/\rho=1/m_{av}$ involves a complicated time
dependent factor $m_{av}$.  Not only does the upper mass cut-off $m_+$
(roughly the main sequence turn-off mass) depend on time, but what is
worse, the distribution of remnants, in the form of black holes, neutron
stars and white dwarfs, does not obey any simple power law, even if
their progenitors did.  In general, therefore, the coefficient $n/\rho$
has to be determined numerically, as a function of time.

\subsubsection{Velocity Dispersions}

In thermal equilibrium, equipartition of kinetic energy tells us how
the velocity dispersions scale for stars with different masses.  We
only have to specify the three-dimensional velocity dispersion $v$ for
one particular mass, say $v(1\msun) = \vsun$, in order to determine
the 3D velocity dispersion $v(m)$ for stars of general mass $m$:
\begin{equation}
  v(m) = \left( {1 \msun \over m} \right)^{1/2} \vsun.
\end{equation}

\subsubsection{Core Radius and Core Mass}

The three choices discussed so far, namely that of an initial mass
function, a density, and a temperature, specify the intensive
thermodynamic properties.  This in turn enables us to calculate the
local rate of collisions, per unit time, and per unit volume.  In
order to extract global information, we have to specify extensive
quantities as well, such as the total volume or total mass of our
system.  This will allow us to determine a global collision rate per
unit time, which we can then compare with that of an astrophysical
system, such as the core of a globular cluster.

For an equal-mass cluster model that is close to thermodynamic
equilibrium, the density drops by roughly a factor three, from the
center to the edge of the core.  This implies that the local density
of collisions, which is proportional to the square of the density,
drops by an order of magnitude.  In the more realistic case of a mass
spectrum the situation is even worse, since the density of the
heavier stars drops off faster than that of the lightest stars.  In
the present paper we will not attempt to model these density dependent
effects, and instead we
will keep the density of all mass groups constant throughout the
region of our simulation.  It is clear, therefore, that our results
are mainly for the purpose of illustration, and that any comparison
with actual systems will have to be taken with many grains of salt.

The only question remaining is the definition of a core radius $r_c$.
For an equal mass system, we have (\cite{spi87})
\begin{equation}
r_c = \left( {3 \over 4 \pi G \rho} \right)^{1/2} v_c.
\end{equation}
In the presence of a mass spectrum, we have to modify this equation.
Although the velocity dispersion is now quite different for different
mass groups, the average kinetic energy per star $(1/N)E_{kin}$ is
independent
of mass, with $N$ the total number of stars in the core.  Rewriting
the above formula, we have:
\begin{equation}
r_c^2 =  {3 \over 4 \pi G \rho} {2\over M} E_{kin},
\end{equation}
where $M$ is the core mass.  For a general mass spectrum, we can
substitute $E_{kin} = (N/2)\msun\vsun^2$ which gives
\begin{equation}
r_c = \left( {3 \over 4 \pi G \rho} \right)^{1/2} 
       \left( {\msun\over m_{av}} \right)^{1/2} \vsun.
\end{equation}
In this expression, the right hand side contains only local
quantities, and the global quantity $r_c$ is given in terms of those.

Note that this is not the only possible generalization of the
equal-mass expression, but it is a natural one, and it reverts to the
original expression in cases where the mass of the core is
dominated by stars in a relatively small mass range, as is the case,
for example, in a post-collapse core of a globular cluster.

Other global quantities can be derived from $r_c$, such as the core
mass $M$:
\begin{equation}
M = {2 \over 3} \pi r_c^3 \rho,
\end{equation}
where we have used the fact that in an isothermal sphere the average
density in the core is roughly half the central density, a
relationship, while not exact, is certainly good enough for our
purpose of relating our results to astrophysical systems.

\subsection{Recipes for Stellar Evolution}
The stars in our computations are evolving. To describe their evolution,
we use the formulae fitted to the results of full stellar evolution
calculations, by \cite*{eft89}. These formulae give the radius
and luminosity (for population I stars)
as a function of time, on the main sequence, in the
Hertzsprung gap, the (sub)giant branch, on the horizontal branch,
and on the asymptotic giant-branch.
We use these population~I recipes, because
the more appropriate data for population~II stars are not available in
the same
convenient form.
In addition to the radius, we need the core mass for stars that have left
the main sequence. We derive these from the luminosity, according to
\cite*{eft89}, and core-mass luminosity relations, according to
\cite*{bs88}, \cite*{pac70} and \cite*{it78}.
The details of this procedure are described in Portegies Zwart and
Verbunt (1996, Section 2.1).\nocite{pzv96}

\subsection{Recipes for Individual Encounters}
In this paper, we only treat single stars, and accordingly the
only outcome allowed for a close encounter is a merged object.
The merging between the two stars in an encounter
in our calculation is generally assumed to
conserve mass, which in fact may be a reasonable approximation
(Benz and Hills 1987, 1989, 1992, Rasio and Shapiro 1991, 1992).
\nocite{bh87}\nocite{bh89}\nocite{bh92}\nocite{rs91}\nocite{rs92}

Only a limited number of simulations of encounters between stars
has been performed, and these does not cover all possible combinations
that may occur in a cluster. Also, different authors do not agree
on the details of the outcomes for the same type of encounter.
We therefore have chosen to use a set of simple prescriptions for
the outcome of stellar collisions, often chosen without detailed
justification.
In the future these prescriptions can be refined, when more accurate
calculations for collisions become available. Meanwhile, our results
will help in determining which of all possible types of encounter
are most frequent, and therefore deserve closer attention.
\begin{table}
\caption[]{Simplified representation of possible merger outcomes.
           The four columns correspond to the four choices given for the
          type of massive star (primary), while the four rows indicate
           the type of less massive star (secondary):
           main-sequence star (ms), (sub)giant (sg), white dwarf (wd) and
          neutron star (ns).
           In this table we do not discriminate between stars
           in the Hertzsprung gap (Hg) or on the first and second ascent
          on the asymptotic-giant branch (AGB).}
\begin{flushleft}
\begin{tabular}{l|cccc}
          & \multicolumn{4}{c}{primary} \\
      star& ms & sg& wd  & ns \\ \hline
          &    &   & wd  & ns \\
       ms & ms & sg& +   & +  \\
          &    &   & disc&disc\\ \hline
          &    &   & wd  & ns \\
       sg & Hg &AGB& +   & +  \\
          &    &   & disc&disc\\ \hline
          &    &   &     &    \\
       wd & sg &AGB& --  & -- \\
          &    &   &     &    \\ \hline
          &    &   &     &    \\
       ns &\TZO&\TZO& -- & -- \\
          &    &   &     &    \\ \hline
\end{tabular}
\end{flushleft}
\label{Tab_mprod}\end{table}

We describe our treatment of the possible outcomes of the encounters
of two stars ordered by the evolutionary state of the more
massive of the two, the primary. Table~\ref{Tab_mprod}
summarizes this treatment.

\subsubsection{Main-sequence primary}
If both stars involved in the encounter are main-sequence stars
the less massive star is accreted conservatively onto the most massive
star.
The resulting star is a rejuvenated
main-sequence star (see Lai et al.\ 1993, Lombardi et al.\ 1995).
\nocite{lrs93}\nocite{lrs95}
The details of this procedure are described in Appendix C4 of Portegies
Zwart and Verbunt (1996).

If the less massive star in the encounter
has a well developed core (giant or subgiant)
this core is treated as the core of the merger product.
The main-sequence star and the envelope of the giant are added together
to form the new envelope of the merger.
In general the mass of the core is relatively small compared
to its envelope
and the star is assumed to continue its evolution through the
Hertzsprung gap.
Note that this type of encounter can only occur when the main-sequence
star is in itself a collision product (e.g.\ a blue straggler).

When a main-sequence star encounters a less massive white dwarf, we
assume that the merger product is a giant, whose core and envelope
have the masses of the white dwarf and the main-sequence star,
respectively.
We then determine the evolutionary state of the merger product, as
follows.
We calculate the total time $t_{\rm agb}$ that a single,
unperturbed star with a mass equal to that of the merged star
spends on the asymptotic giant-branch,
and the mass $m_{\rm c,agb}$ of its core at the tip of the giant branch.
The age of the merger product is then calculated by adding
$t_{\rm agb}m_c/m_{\rm c,agb}$ to the age of an unperturbed star
with the same mass at the bottom of the asymptotic giant branch.
For example, a single, unperturbed 1.4$\msun$ star leaves the
main-sequence
after 2.52$\,$Gyr, spends 60$\,$Myr in the Hertzsprung gap,
moves to the horizontal branch at $2.96\,$Gyr, and reaches the tip
of the asymptotic giant branch after $3.06\,$Gyr, with a core of
$0.64\,\msun$.
Thus, if a 0.6$\,\msun$ white dwarf mergers with an $0.8\,\msun$
main-sequence star, the merger product has an age of 2.87$\,$Gyr,
leaving it another 180$\,$Myr before it reaches the tip of the asymptotic
giant-branch.

If the less massive star is a neutron star
a Thorne \Zytkow~object (\cite{tz77}) is formed.

\subsubsection{Evolved primary}
When a (sub)giant or asymptotic branch giant
encounters a less massive main-sequence star, the
main-sequence star is added to the envelope of the giant, which
stays in the same evolutionary state, i.e.
remains a (sub)giant, c.q. asymptotic branch giant.
Its age within that state is changed, however, according to the
rejuvenation calculation described in Section C.3 of Portegies Zwart
and Verbunt (1996).
For example, an encounter of a giant of $0.95\,\msun$ and age
$11.34\,$Gyr with a 0.45$\,\msun$ main-sequence star produces
a giant of 1.4$\,\msun$ with an age of $2.67\,$Gyr.

When both stars are (sub)giants the two cores are added together and form
the core of the merger product
(see also the results of the smoothed particle
hydrodynamics computations performed by Davies et al.\ 1991 and
Rasio and Shapiro 1995).\nocite{dbh91}\nocite{rs95}
Half the envelope mass of the (less massive) encountering star
is accreted onto the primary.
The merger product continues its evolution starting at the next
evolutionary state;
thus a (sub)giant continues its evolution on the horizontal branch and a
horizontal branch star becomes a asymptotic-giant branch star.
The reasoning behind this assumption is that an increased core mass
corresponds to a later evolutionary stage.

If the less massive star is a white dwarf
then its mass is simply added to the core mass of the giant,
and the envelope is retained.
If the age of the giant before the encounter exceeds the total life
time of a single unperturbed star with the mass of the merger,
then the newly formed giant immediately sheds its envelope,
and its core turns into a single white dwarf;
if not then the merged giant is assumed to have the same age (in years)
as the giant before the collision, and continues its evolution as a
single unperturbed star.

If the encountering star is a less massive neutron-star
a Thorne \Zytkow~object is formed.

\subsubsection{White-dwarf primary}
In an encounter between a white dwarf and a less massive main-sequence
star,
the latter is completely disrupted and forms a disk around the white
dwarf (\cite{rm90}, \cite{rs91}). The white dwarf accretes from this disk
at a rate of one percent of the Eddington limit. 
If the mass in the disc exceeds 5\% of the mass of the white dwarf, the
excess mass is expelled from the disc at a rate equal to the Eddington
limit.

If a white dwarf encounters a less massive (sub)giant, a new white
dwarf is formed with a mass equal to the sum of the pre-encounter
core of the (sub)giant and the white-dwarf.
The newly formed white dwarf is surrounded by a disk formed from
half the envelope of the (sub)giant before the encounter.
If the mass of the white dwarf surpasses
the Chandrasekhar limit, it is destroyed, without leaving a remnant
(\cite{nk91} and \cite{lt95}).

Collisions between white dwarfs are ignored.

\subsubsection{Neutron-star or black-hole primary}
All encounters with a neutron star or black hole primary
lead to the formation of a massive
disk around the compact star.
If the compact star had a disk prior to the collision, this disk is
expelled.
This disk accretes onto the compact star at a rate of 5\% of the
Eddington limit.
An accreting neutron star
turns into a millisecond radio-pulsar, or -- when its mass
exceeds $2\msun$ -- into a black hole.
Mutual encounters between neutron stars and black holes are ignored,
as are collisions between these stars and white dwarfs.

\subsection{Monte Carlo Simulations of Ensembles of Encounters}
Each star in our model can encounter any of the other stars.
To reduce computational cost, we bin the stars in intervals
of mass and radius, and compute the probability for encounters
between bins, giving all stars in one bin the same mass and radius,
and, through Eq.~2, choosing their velocities from the same distribution.
The cross section $\sigma_{ij}$
for a encounter with a distance of closest approach
within $d$ between a star from bin $i$
and a star from bin $j$ contains a geometrical and a
gravitational focusing contribution:
\begin{equation}
        \sigma_{ij} = \pi d^2 \left( 1 + 2G \frac{m_i + m_j}
                                                  {v_{ij}^2 d}
                               \right),
\label{sigma}\end{equation}
where $v_{ij}$ is the relative velocity between the stars at infinity.
For the minimum separation between the two stars that
leads to a collision $d = 2(r_i + r_j)$ is used.
(The exact distance at which the transition between merger and binary
formation occurs is not known --\cite{koc92}, \cite{lrs93}--,
we choose the factor 2 arbitrarily.)

  In the present paper, we model the stellar distributions as being
  spatially homogeneous.  In order to make contact with astrophysical
  applications, we will consider our stars to be contained within in a
  fixed sphere with radius $r_c$.  While we can consider this radius to
  stand for the notion of `core radius' in a post-collapse cluster, we
  want to point out that this interpretation is only an approximate one.
  In realistic star clusters, there is a significant drop in density
  across the core, from the center to the core radius.  For most stars
  the density drops by roughly a factor of three, but for the heavier
  stars, such as neutron stars and especially black holes, this factor
  can be much larger.

  The encounter rate $\Gamma_{ij}$ of stars from bin $i$ with stars
  from bin $j$, anywhere in the volume of the sphere with radius $r_c$
  is given by two separate equations:
  \begin{equation}
  \Gamma_{ij} = 
                 \left\{ 
                        \begin{array}{crr}
                  {1\over 2}n_i(n_j-1)  & \langle \sigma_{ij}v_{ij}
                                          \rangle {4\over 3}\pi r_c^3
                &  \,  \mbox{for $i = j$} \\
                  n_in_j                & \langle \sigma_{ij}v_{ij}
                                          \rangle {4\over 3}\pi r_c^3
                &  \,  \mbox{$i<j$},
                        \end{array}
                 \right.
\end{equation}
  where $n_i$ and $n_j$ are the number densities of stars in bins $i$
  and $j$, respectively, and where $\langle\, \rangle$
  indicates averaging over the 
  the distribution of relative velocities $v_{ij}$ (Note that we
  should have written the last equation with an extra factor
  $1/2$, if we would have summed over all combinations $i \neq j$,
  in order to avoid double-counting of collisions).

Since the stars in bins $i,j$ have Maxwellian velocity distributions
with root-mean-square velocity $v_i$ and $v_j$, given by Eq.(2), the
relative
velocities $v_{ij}$ also have a Maxwellian distribution,
with a root-mean-square velocity given by $\sqrt{{v_i}^2+{v_j}^2}$. Hence

\begin{eqnarray}
        \langle \sigma_{ij} v_{ij} \rangle &=& \frac{4l^3}{\sqrt{\pi}}
                        \int_0^\infty v_{ij}^3\sigma_{ij}
                        \exp (-l^2v_{ij}^2) dv_{ij} \nonumber \\
                      & = & {2\pi d^2\over l\sqrt{\pi}}
                            \left( 1+2G{m_i+m_j\over d}l^2 \right),
\label{sigmav}\end{eqnarray}
where we have defined
\begin{equation}
        l^2 = {3\over 2({v_i}^2+{v_j}^2)}.
\end{equation}

With this result, we write the encounter rates $\Gamma_{ij}$ in
convenient units:
\begin{eqnarray}
        \Gamma_{ij}
                     &=&        \frac{n_i}{10^3 \mbox{pc}^{-3}}
                                \frac{n_j}{10^3 \mbox{pc}^{-3}}
                          \left(\frac{r_{\rm c}}{\mbox{pc}}\right)^3
                                            \nonumber \\
                     &\times&
                        [
                        3.61 \cdot 10^{-3}
                              \left(\frac{m_i+m_j}{\msun}\right)
                                            \left(\frac{r_{ij}}{\rsun} \right)
                                            \left(\frac{\Kms}{v_{ij}} \right)
                                            \nonumber \\
                     &+& 6.31 \cdot 10^{-9}
                             \left(\frac{r_{ij}}{\rsun}\right)^2
                             \left(\frac{\Kms}{v_{ij}}\right)^{-1}
                        ]
                                            \nonumber \\
                     & &                   \,[\mbox{Myr}^{-1}].
\label{rate}\end{eqnarray}
The total encounter-rate follows as
\begin{equation}
        \Gamma = 
                 \sum_{ij =1}^{N} \Gamma_{ij} \equiv {1\over \tau_{\rm enc}},
\label{renc}\end{equation}
where $N$ gives the total number of bins in mass and radius and
$\tau_{\rm enc}$ is the average time interval between two encounters.

The stellar population in our calculation changes both due to encounters
between stars, and due to evolution of the stars.
The shortest evolutionary time-scale of importance to us is the time
scale
on which the evolving stars expand; the fastest evolving star in the
sample is used to set the evolution time scale
\begin{equation}
\tau_{\rm ev}\equiv \min (R/{\dot R}),
\end{equation}
where $R$ and $\dot R$ are the stellar radius and its time derivative,
respectively.

At the beginning of each time step, we distribute the stars over the bins
in radius and mass, calculate the number densities of stars in each
bin, and the
evolution and collision time scales $\tau_{\rm ev}$ and $\tau_{\rm enc}$.
The sum over all bins $ij$ is less daunting as may appear at first sight,
as many bins contain no stars. This is illustrated in
Figure~\ref{fig_rencf_C}. The time step to be taken is then calculated as
\begin{equation}
dt = \min (0.2\tau_{\rm enc},\tau_{\rm ev}),
\end{equation}
to ensure that changes in the stellar population are followed with
sufficient resolution.

At this point, a rejection technique is used to keep track of
collisions, as follows.
We choose a random number between 0 and 1. If this number is larger
than $\Gamma dt$, we
conclude that no collision has occurred. We evolve all stars
over a time $dt$, and continue with the next step.

If the random number is smaller than $\Gamma dt$, a collision has
occurred.
In calculating the sum (Eq.~\ref{renc}) over the bins,
we keep track of the partial
sum after addition of each bin combination $ij$.
The first bin combination for which this growing partial
sum exceeds the random number
identifies the bins involved in the collision.
We then assign a sequence number to each star in bin $i$, and select
one of these numbers randomly; and repeat this for bin $j$. If $i$ and
$j$ are identical, care is taken that the same star is not selected
twice.
From the prescriptions in the previous section, we decide the outcome
of the collision between the two selected stars.

We then select another random number between 0 and 1, to see whether
a second collision has occurred. If so, we determine its outcome.
This procedure is repeated until a random number
larger than $\Gamma dt$ is found,
which indicates that no further encounter has occurred in the time step
under
consideration.

After each time step $dt$ a number of stars equal to the number of
encounters that have taken place is lost from the
stellar system; these stars have merged into single objects.
For each lost star
a new star is added to the computation, in order to guarantee a
constant number of stars.
The mass of this `halo guest' is determined by the present-day
mass-function of the cluster.

\begin{figure}
\hspace*{1.cm}
\psfig{figure=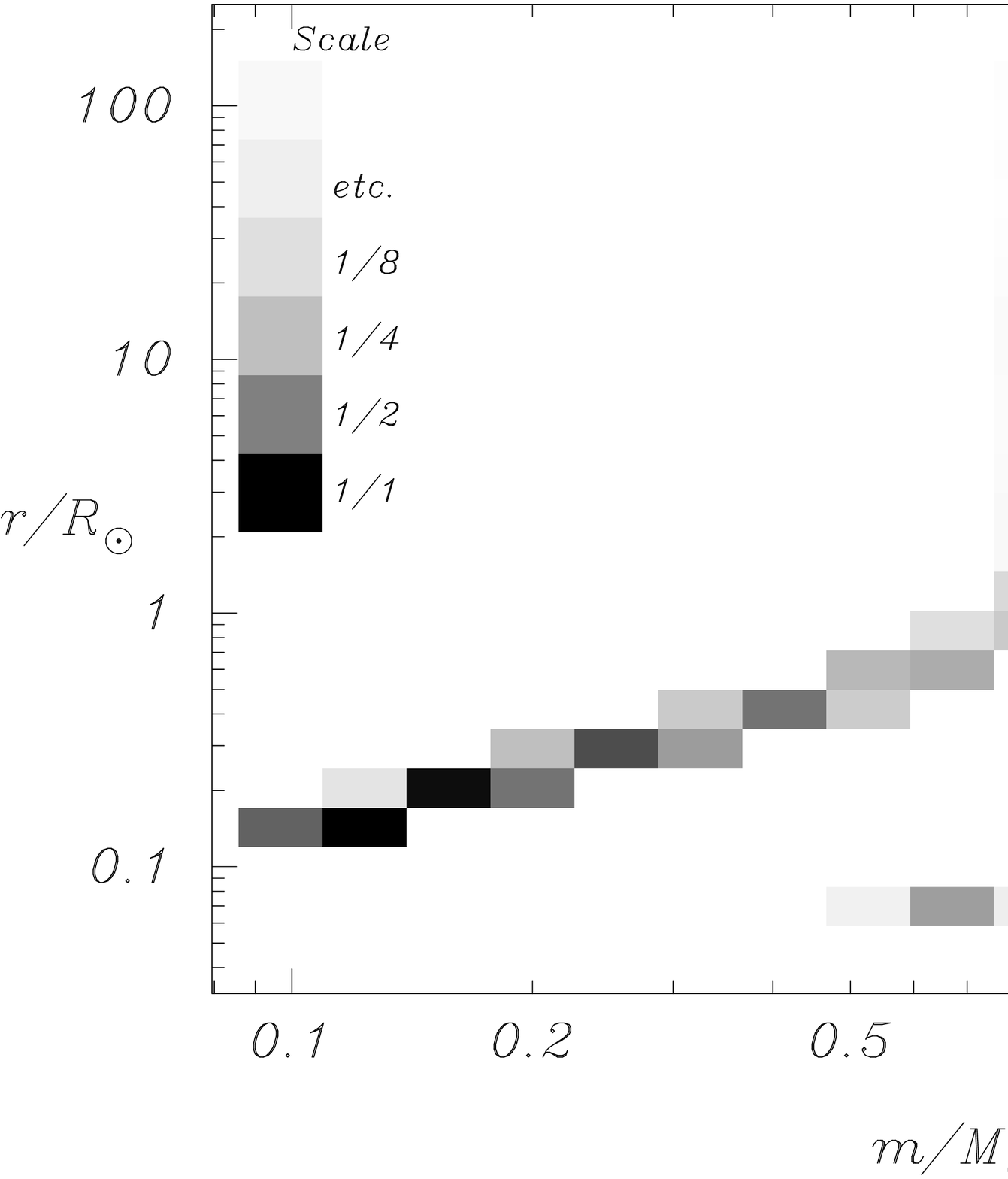,width=4.0cm}
\caption[]{Relative encounter probabilities in model calculation $S$, at
time $t=12\,$Gyr, when the turnoff mass is $M_{\rm to}=0.91\,\msun$,
for a single star with $1\msun$ and $1\rsun$
as a function of mass and radius of the other star
involved in the encounter.
Darker shades indicate higher probabilities.
The compact stars (nominally with zero radius) are shown
as a bar below 0.1$\,\rsun$: neutron stars between 1.34 and 2
$\msun$ and  white dwarfs at lower masses.
All other stars with radius in excess of the radius at the turnoff are
the evolved stars. The masses of these stars is similar to the turnoff
mass.
A small fraction of blue stragglers is visible as an extension of
the main-sequence (to the right of the turnoff).

The vertical bar in the upper left corner presents a scaling to the gray
shades.
The lowest square corresponds to an encounter rate 
of once every 12.7~Tyr decreasing with a factor of two for 
each subsequent square.
The integrated encounter frequency of the $1\msun, 1\rsun$ star is
1 encounter every 1.68~Tyr.
Almost 13\%\ of the encounters occur with a main-sequence star with
a mass of about $\sim 0.13 \msun$
(black squares below and to the left).
}
\label{fig_rencf_S}\end{figure}

\section{A Dynamically Evolving Salpeter Mass Function}

\begin{figure*}
\hspace*{4.5cm}
\epsfxsize = 4.0cm
\epsffile{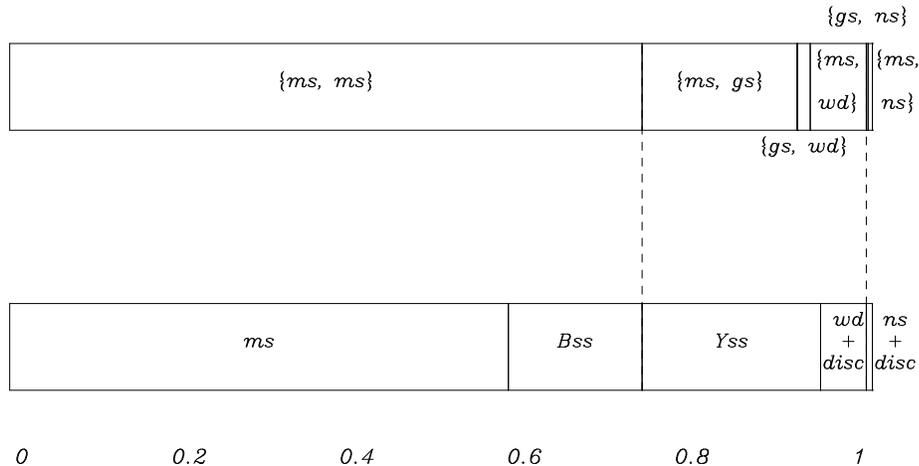}
\caption[]{Relative frequencies of various types of encounters 
(upper panel, the curly brackets indicate the collision)
and their outcomes (lower panel), for model computation $S$,
integrated over the duration of the calculation.
Abbreviations: ms main-sequence star, gs giant, wd white dwarf,
ns neutron star, Bss blue straggler, Yss yellow straggler.}
\label{fig_rencprob}\end{figure*}

A total of two models are computed, one  with a Salpeter type mass function
which we call model $S$ (from Salpeter), and one model which we call 
model~$C$ (from Collapsed) with a mass function that is affected strongly by
mass segregation.

In the volume of the stellar system in model $S$, we sprinkle stars according
to a Salpeter mass distribution between 0.1\msun and 100\msun.  
The total number of stars is
irrelevant, since we are considering these stars to be contained in a
laboratory-type enclosure, with a thermal distribution of stellar
velocities.  Our choice for the `temperature' of this distribution is
fixed by requiring that stars with a mass $m=1\msun$ will have a
one-dimensional velocity dispersion of 10.0~km/s, in conformity with
the same choice made in \S5.
The radius of the core was chosen to be $r_c=4$~pc
and the computation is started at $t=0$ and terminated at an age of 16\,Gyr.
For the computation of the encounter rate a total of
30 bins in mass, equally spaced in the logarithm of the mass
between 0.1 and 100 $\msun$, and 30 bins in radius, equally spaced
in the logarithm of radius between 0.1 and 2000 $\rsun$ are used.
An additional bin with zero radius is used for the compact stars,
i.e. the white dwarfs, neutron stars and black holes.

\begin{table*}
\caption[]{Parameters of the different model computations,
and corresponding characteristics.
Subsequent columns give the name of the model,
indication whether a Salpeter mass function or a mass function that is
affected by mass segregation is used, core radius, 
3-dimensional velocity dispersion $\vsun$ time at which
encounters are started,  
the age of the population at the start of the dynamical interactions, 
the central stellar number-density in the core,
the ratio of the number of stars in the computation
to the actual number of stars in the core,
the number of encounters during the calculation per star, and
the average time between two encounters, anywhere in the core.}
\begin{flushleft}
\begin{tabular}{llcr|rllrl}
Model &mf&$r_{\rm c}$&$\vsun$&$t_{\rm cc}$
      &$\log n_c$&$f_{\rm c}$
      & $n_{\rm enc}$ & $\tau_{\rm enc}$ \\ \hline
         &&[pc] &[km/s]&[Gyr]&[$\star$~pc$^{-3}$] &&$\star^{-1}$&[Myr]
        \\  \hline
$S$&Salpeter  &4.0 &17.3& 0&3.93 & 0.298 &0.002& 5.17 \\
$C$&Segregated&0.1 &17.3&10&6.64 & 8.750 &0.660& 1.28 \\ \hline 
\end{tabular}
\end{flushleft}
\label{Tab_init}\end{table*}

Figure~\ref{fig_rencf_S} shows for model computation $S$,
the relative probabilities of encounters
with various types of stars for a $1\msun, 1\rsun$ star, at an age
of the cluster of 12$\,$Gyr. 
Due to the small encounter frequency hardly any collision products are
present in the stellar system. Only a small number of blue
stragglers (stars with a mass larger than the turnoff and with
similar radii) have finite probability to be involved in an encounter.
The most probable partners for an encounter with a $1\msun, 1\rsun$ star
are the stars at the low end of the main sequence. 

In Figure~\ref{fig_rencprob} we show the relative frequencies of
encounters
of different types, and of the resulting collision products for model $S$.
Because the steep mass function the collisions rate is 
dominated by main-sequence stars; the fraction of collisions involving 
giants is only small.
The most frequent type of encounter is one involving two main-sequence stars,
leading to a main-sequence merger remnant with a mass smaller than the turnoff
mass or a blue straggler when the mass of the merger exceeds the turnoff mass. 
If the mass of the merger is less than the
turnoff mass, the product is a main-sequence star which is younger
than primordial main-sequence stars with the same mass. Such a
star will be left behind as a blue straggler once the primordial
main-sequence stars leave the main-sequence.
Yellow stragglers, i.e. giants not on the main (sub)giant branch of the
cluster (which approximately coincides with the
evolutionary track of a star with the turn off mass),
can be formed directly from encounters between
a main-sequence star and a giant, between a main-sequence star and a
white dwarf and between a giant and a white
dwarf, in decreasing order of importance;
encounters between two giants are extremely rare.
Our prescriptions put every merger product on the
evolutionary track of an ordinary star; the presence of yellow
stragglers in our calculations is
therefore only due to the formation of giants with a mass
larger than the turnoff mass. 

\section{A More Realistic Mass Function}
The initial conditions for the mass function of the computation of model
$C$ (for collapsed cluster core) are chosen to be more realistic, in the sense
that the mass function is flattened due to mass segregation in the
previous evolution of the stellar system. 
The lack of detailed computations concerning the present-day mass function in
the cores of globular clusters, justifies our choice 
to use a mass function similar to the one described by \cite*{vm88}.
For the mass function of model $C$
we consider three classes of objects: non-degenerate stars 
(main-sequence stars and giants), white dwarfs, and neutron stars.
The more massive stars have all evolved, and left inert remnants
(white dwarfs or neutron stars).
We assign a certain fraction of the
total number of stars in the stellar system to each of these classes.
All neutron stars (5\% of the total number of stars) are assumed
to have the same mass (of $1.34\msun$).
The mass distribution within the two other classes are described
with power-laws with a slope of $\alpha =0$ for the main-sequence stars 
and the (sub)giants and a slope of $\alpha =1$ for the white dwarf
progenitors.
At the start of the dynamical modeling a total number fraction of 
main-sequence stars and giants of 70\% is chosen,
this number decreases as the stellar system evolves.
The minimum initial mass of a main-sequence star is chosen to be
0.2\msun\ instead of the 0.1\msun\ for models $S$.

The numbers of stars in the different classes change
as time evolves due to stellar evolution, encounters between stars,
and due to the addition of a star, each time that
the number of stars has decreased by one in a merger process.

\begin{figure}
\hspace*{1.cm}
\epsfysize = 5.5cm
\epsffile{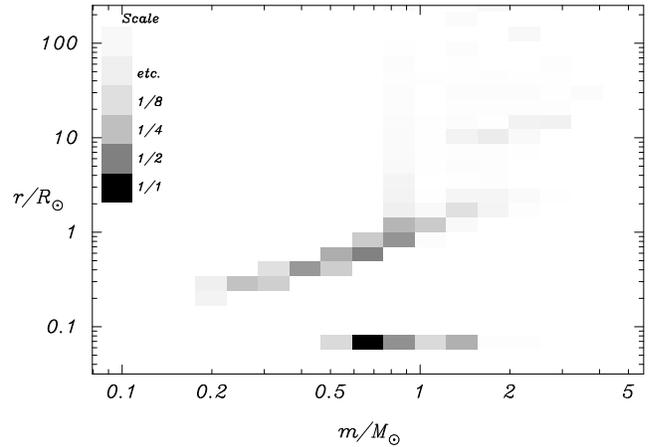}
\caption[]{Relative encounter probabilities in model calculation $C$, at
time $t=12\,$Gyr, when the turnoff mass is $M_{\rm to}=0.91\,\msun$,
for a single star with $1\msun$ and $1\rsun$
as a function of mass and radius of the other star
involved in the encounter (similar to Fig.~\ref{fig_rencf_C}).
The high encounter-rate and different mass function result in an
enormous enrichment of collision products in the stellar system.
Besides the small fraction of black holes (nominally with zero radius an
with a mass larger than 2\msun), there is also a rich population of 
blue stragglers
(in the area with a mass larger than the turn off and a radius larger than
about one \rsun) and yellow stragglers (stars with a radius larger than
that of the blue stragglers).
Except for the neutron stars and black holes (nominally with zero radius
an with a mass larger than 2\msun)
all stars with mass in excess of the turnoff mass are
the products of previous encounters.

The vertical bar in the upper left corner presents a scaling to the gray
shades.
The lowest square corresponds to an encounter rate 
of once every 21~Gyr decreasing with a factor of two for each subsequent square.
The integrated encounter frequency of the $1\msun, 1\rsun$ star is
1 encounter every 3.1~Gyr.
Almost 15\%\ of the encounters occur with a white dwarf with
a mass of about $\sim 0.7 \msun$
(black square in the middle and below).
}
\label{fig_rencf_C}\end{figure}

\begin{figure*}
\hspace*{4.5cm}
\psfig{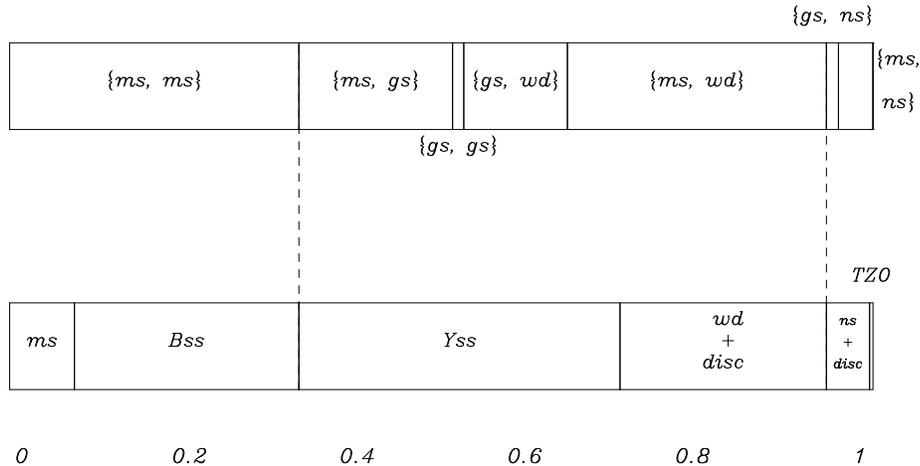}
\caption[]{Relative frequencies of various types of encounters (upper
panel)
and their outcomes (lower panel), for model computation $C$,
integrated over the duration of the calculation 
(see also Fig.~\ref{fig_rencf_S}).

Apart from the variation in the relative encounter frequencies 
between various types of stars, encounters
between two giants become noticeable and Thorne \Zytkow\ objects (\TZO) 
appear (in very small numbers) as the result of a collision.
}
\label{fig_rencprobC}\end{figure*}

Model~$C$ has a core radius of
$r_{\rm c} = 0.1$~pc and a 1-dimensional velocity-dispersion
for a 1~\msun star of 10~km/s.
We switch-on the dynamics at $t_{cc}=$10~Gyr and terminate the model
at $t=$16~Gyr.

The number of stars used in the computation is higher than
the calculated number of stars in the core for the parameters of model
$C$; as a result the Poissonian noise in our calculation is smaller
than it would be in an actual core.

Figure~\ref{fig_rencf_C} shows for model computation $C$,
the relative probabilities of encounters
with various types of stars for a single $1\msun, 1\rsun$ star, at an age
of the cluster of 12$\,$Gyr. At this age, products of previous
encounters are already present in the cluster, and have a finite
probability of undergoing another encounter. However, the
most probable partner for an encounter with a $1\msun, 1\rsun$ star
is a white dwarf with a mass of about 0.7\msun.

The relative importance of the various types of encounters
is very different in model $C$ compared to model $S$, as illustrated in
Figure~\ref{fig_rencprobC},
and consequently the relative frequencies of merger
outcomes are very different as well.
The fraction of collisions that directly result in the formation of a
blue straggler rises sharply as does the relative formation-rate of yellow
stragglers and white dwarfs with a massive disc. 
Because the  mass function in model $C$ is flat, the region of the main
sequence around the turn-off
is well populated with massive main-sequence stars and consequently the
total number of giants is much larger than in model $S$ where a steep
mass function is used.

\subsection{An Evolved H-R Diagram}
\begin{figure}
\hspace*{1.cm}
\psfig{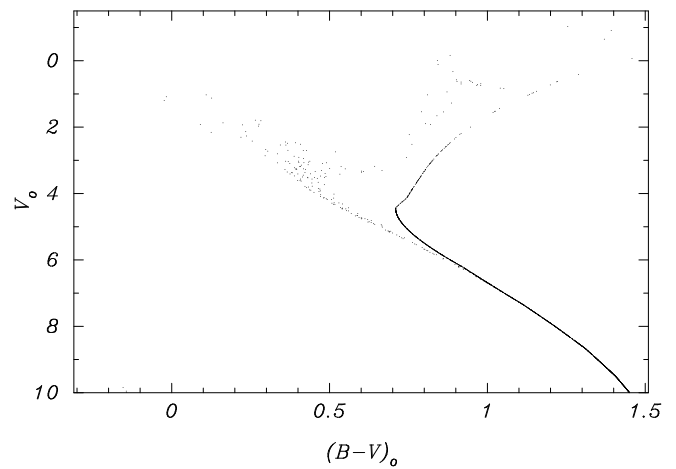}
\caption[]{Hertzsprung-Russell diagram of model $C$, at ca.\ $t=12$~Gyr.
           $10^4$ stars (corresponding to about the
           total number of stars in the core) were selected randomly from all
           stars involved in the simulation.}
\label{fig_hrd_C12}\end{figure}

\begin{figure}
\hspace*{1.cm}
\psfig{figure=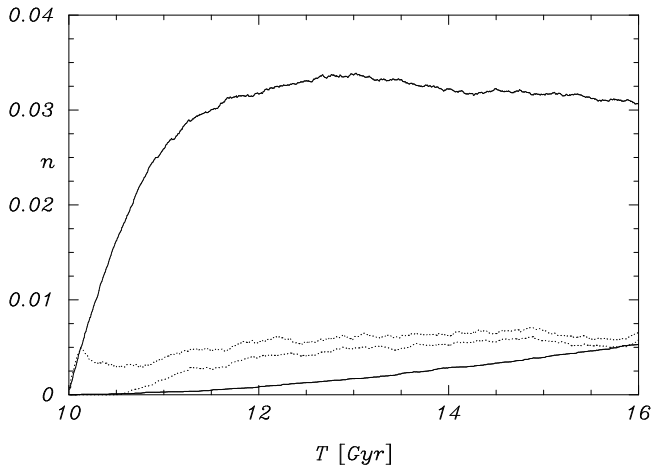,width=4.0cm}
\caption[]{Fraction of stars in the computation of model $C$
that are blue stragglers (upper solid line) and the fraction
of stars on the main-sequence that were left behind as blue stragglers
when primordial stars of equal mass evolved into giants
(lower solid line), as a function of time. Due to the slow evolution
on the main sequence, the lower line is less susceptible to
Poissonian fluctuations.
The dotted lines show the fraction of stars that are yellow stragglers,
for all yellow stragglers (upper dotted line) and for those that evolved
from blue stragglers (lower dotted line).}
\label{fig_nbgss}\end{figure}

A Hertzsprung--Russell diagram of model $C$ after about 12$\,$Gyr
is shown in Fig.~\ref{fig_hrd_C12}.
The dots (representing individual stars) that are positioned in the color
magnitude diagram at a position that deviates from the isochrone 
of the stellar system are the result of a collision.
Blue stragglers can be identified close to the zero-age main-sequence but
are bluer and more luminous than the turn-off, whereas yellow stragglers are
situated above the giant branch.
Because the stars in our calculation evolve, the number of
collision products present at any time in the core is not
at all proportional to their formation rate. For example,
blue stragglers (a main sequence star with mass $M > M_{\rm to}$), 
formed by merging
of two main-sequence stars, often evolve into giants
before our calculation is stopped, because of the short
main-sequence lifetime of more massive stars.
Evolving blue stragglers turn into yellow stragglers, and
in fact most of the yellow stragglers present in the cluster have
evolved from blue stragglers. The yellow stragglers formed
directly from collisions with giants evolve too fast to
contribute as strongly to the presence of yellow stragglers.
This is illustrated in Figure~\ref{fig_nbgss}, which also shows that the
fraction of stars that are yellow stragglers is rather constant
throughout the computation.

Merged main-sequence stars with a mass smaller than the turnoff
mass upon formation are left behind as blue stragglers when the
equally massive primordial stars leave the main sequence.
As illustrated in Figure~\ref{fig_nbgss} (lower solid line), 
the fraction of such blue stragglers is relatively small.
On the other hand, the fraction of stars that are blue
stragglers rises rapidly at first, but levels off when the
evolution rate of blue stragglers into yellow stragglers and
beyond becomes competitive with their formation rate.
Thus, the fraction of stars that are blue stragglers does not
rise much above 3\% at any given time, even though 26\% of the
stars in the computation is directly turned into a blue straggler
at some time or after after a collision.
The dotted line in Figure~\ref{fig_nbgss} illustrates that the total
number of yellow stragglers is roughly constant from the beginning of the
dynamical simulation.

\begin{figure}
\hspace*{1.cm}
\psfig{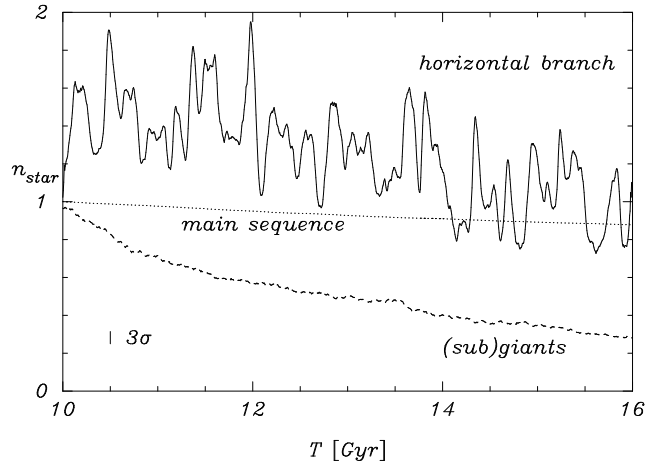}
\caption[]{Total number of stars formed in model $C$ during the computation with
           dynamical encounters divided by the number formed from
           a non-dynamical model as a function of time.
           The dotted line indicates the fraction of main-sequence stars,
           the dashed line the stars on the Hertzsprung gap 
           and (sub)giant branch, and the
           solid line the relative fraction of
           horizontal-branch stars (averaged in 100~Myr intervals).

           Stars on the (sub)giant branch are
           more depleted than the main-sequence stars as time evolves.
           The fraction of horizontal-branch stars is roughly twice as
           large in the stellar system where collisions are included.
           The Poissonian noise for the main-sequence stars is
           smallest, as expected, 
           followed by that in the number of (sub)giants.
           The noise in the fraction of horizontal branch stars is largest.
           The $3\sigma$ error bar (lower left) indicates the Poissonian
           error for the giants (the dashed line).
}\label{depletion}\end{figure}

Giants which undergo a collision become more massive in our
prescription, and thus evolve faster than their unperturbed
counterparts. As a result, the number of giants in the model
is smaller than it would have been in a cluster without collisions,
as illustrated in Figure~\ref{depletion}.
At the end of the computation the number of giants is depleted by
roughly 70\%. The fraction of stars on the horizontal branch is roughly
twice that expected from a non-dynamically evolving stellar
system. 
This enhancement of the fraction of horizontal branch stars is the result
of two effects: 
most collisions between a giant and another star result in aging of the
giant which is then evolved closer towards the horizontal branch and
the majority of the collisions
between a main-sequence star and a white dwarf results in the formation
of a star that is about to terminate its giant lifetime 
(e.g.\ close to or on the horizontal branch).

\section{Conclusions}
The models discussed in this paper are very crude in their
treatment of the encounter processes, of the result of a collision
between
two stars, and of the evolution of the merger products.
Apart from these approximations and the fact that we use a stellar evolution
model for population~I instead of pop.~II stars,
the adopted mass function is also highly uncertain.
Nonetheless, some interesting results can be delineated.

Comparison with the calculations of \cite*{db95}
shows the effect of allowing  the merger products to
evolve. An immediate consequence of this is the lower
prediction for the number of blue and yellow stragglers
present in the cluster (as is clear from Figure~\ref{fig_nbgss}).
The formation rates of blue and yellow stragglers give a poor indication
for the actual number of stragglers present in the cluster at a
particular instant.

Due to the low density of model $S$ the collision frequency is small.
The steep Salpeter mass function also suppresses the encounter rate and
the production of stellar curiosities; the majority of the collisions
involve two rather low mass main-sequence stars which results in a
merger that evolves too slow to produce a blue stragglers within the
time span of the simulation.

The Hertzsprung-Russell diagram of our model
cluster (model $C$) shows that blue stragglers close to the turn-off point
lie on the main sequence, whereas blue stragglers above the
turnoff point are mostly found at some distance from the main sequence.
The reason for this is that collisions only become important
in the cluster when an initial period of low density is
followed by the contraction of the cluster core.
The more massive blue stragglers are formed in collisions
between stars close to the terminal-age main sequence, and
evolve relatively quickly.
Blue stragglers close to the turn-off are formed in collisions between
relatively low-mass stars which did not evolve very far away from
the zero-age main sequence, and therefore also the merger products are
close to the zero-age main sequence, and evolve slowly.
Thus, the point where blue stragglers have left the
main sequence gives an indication of the time when collisions in the
cluster became frequent (see also \cite{pz96a}).

Our model $C$ predicts a depletion of giants, in the core only,
up to $\sim 50\%$ shortly after $t_{\rm cc}$
relative to a collision-less stellar system, in
globular clusters with a collapsed core where
the fraction of horizontal-branch stars is enhanced.
Consequently the depletion of giants relative to the number of horizontal
branch stars is strongly present in the high-density stellar system.
Collisions between single stars
cannot explain the observation that giants can be
depleted well outside the core or completely absent in it,
as observed in the core of M~15 (\cite{dpc+91}).

In our simulated cluster cores the total number of white dwarfs that
exceed the Chandrasekhar limit due to accretion from a circum-stellar
disc is small, even in the cluster simulation with the highest
density. In model $C$ 8\%\ of the white dwarfs experience an
accretion-induced collapse, which (after correction for the ratio
$f_{\rm c}$ between the number of stars in the model and in an actual
core -- see Table~\ref{Tab_init}) corresponds to 190 supernovae of
type Ia during the 6$\,$Gyr of our calculation.
If all of these collapses would lead to the formation of a
neutron star, and if all of these would remain in the core,
this would be a substantial addition to the total number of neutron
stars in the core, which is about 460 (after correction for
$f_{\rm c}$) at the start of our calculation.
This result, however, strongly depends on the adopted mass function for
the white dwarfs.
The formation-rate of neutron stars with an accretion disc and the
subsequent formation of a  recycled pulsar or black hole is (to first
order)
linearly dependent on the number of neutron stars, which depends not
only on the initial mass function but also on the subsequent mass
segregation
in the cluster.

The encounter rates between neutron stars and main-sequence
stars are similar in our calculations to the rates found in
the calculations by \cite*{vm88},
by Di~Stefano \& Rappaport (1992) and by Davies \&\ Benz (1995).
\nocite{vm88}\nocite{db95}\nocite{dr92}
After a collision between a neutron star and another cluster member the
merged object becomes visible as an X-ray source (for at most 1~Gyr)
after which it becomes a recycled pulsar or, if its mass exceeds 2~\msun,
a black hole.
The total number of such X-ray sources, recycled pulsars or black holes
scales linearly, in first order, with the number of neutron stars in the
cluster core, which is rather uncertain.

Our computations reveal that collisions between single stars
result in a small number of recycled pulsars: about 70 are formed
in a core (after correction for $f_{\rm c}$) according to model $C$.
Whether this is enough to explain the observed numbers is not
clear. The intrinsic luminosity distribution of millisecond
pulsars, and hence the fraction of them that is detectable
in a typical cluster, is not known; and some clusters with high
encounter rates show remarkably few recycled pulsars, the
globular cluster NGC~6342 is an example (see \cite{lyn93}).

\acknowledgements{This work was supported in part by the Netherlands
Organization for Scientific Research (NWO) under grant PGS 78-277.
SPZ thanks the Institute for Advanced Study for the 
hospitality extended during his visits.}

\bibliographystyle{aabib}
\bibliography{sdyn}
\end{document}